\documentclass[aps,pra,twocolumn,showpacs]{revtex4-1}
\usepackage{graphicx,color}
\usepackage{amsfonts,amsmath, amsthm, amssymb,graphicx,hyperref}
\usepackage{xcolor}
\usepackage{graphicx}
\usepackage[matrix,frame,arrow]{xy}
\usepackage{pgf}
\begin{document}
%%%%%%%%%%%%%%%%%%%%%%%%%%%%%%%%%%%
%
\title{Majorana representation of adiabatic and superadiabatic processes in three-level systems}
%%%%%%%%%%%%%
\author{Shruti Dogra}
\email{shruti.dogra@aalto.fi}
\affiliation{QTF Centre of Excellence, Department of Applied Physics, Aalto University School of Science, P.O. Box 15100, FI-00076 AALTO, Finland}

\author{Antti Veps\"{a}l\"{a}inen}
\affiliation{QTF Centre of Excellence, Department of Applied Physics, Aalto University School of Science, P.O. Box 15100, FI-00076 AALTO, Finland}

\author{G. S. Paraoanu}
\affiliation{QTF Centre of Excellence, Department of Applied Physics, Aalto University School of Science, P.O. Box 15100, FI-00076 AALTO, Finland}
%%%%%%%%%%%%%
\begin{abstract}
We show that stimulated Raman adiabatic passage (STIRAP) and its superadiabatic version (saSTIRAP) have a natural geometric two-star representation on the Majorana sphere. In the case of STIRAP, we find that the evolution is confined to a vertical plane. A faster evolution can be achieved in the saSTIRAP protocol, which employs a counterdiabatic Hamiltonian to nullify the non-adiabatic excitations. We derive this Hamiltonian in the Majorana picture and we 
observe how, under realistic experimental parameters, the counterdiabatic term corrects the trajectory of the Majorana stars toward the dark state. We also introduce a spin-1 average vector and present its evolution during the two processes, demonstrating that it provides a measure of non-adiabaticity. We show that the Majorana representation can be used as a sensitive tool for the detection of process errors due to ac Stark shifts and non-adiabatic transitions. Finally, we provide an extension of these results to mixed states and processes with decoherence.

\end{abstract}
%%%
\maketitle
%%%
\section{Introduction}
\par

Geometric representations play a key role in modern quantum information science. As far as the dynamics of quantum states is concerned, they may be used as a probe to look into various processes and develop intuitive ideas. For a spin-1/2, the mapping of states to the two-dimensional Bloch sphere is a well-known result dating back to the early work of Bloch and Rabi \cite{BlochRabi1945}, a representation that is ubiquitously used nowadays for visualizing the states of qubits~\cite{nielsen-book-02}. Clearly, for higher-dimensional Hilbert spaces this is a non-trivial task~\cite{bengtsson-book-06}. 
Majorana's key insight from 1932~\cite{majorana-nc-1932} was to represent these states as a ``constellation'' of several points (Majorana stars). 
According to this idea, a pure 
state of a particle with spin `$j$' is represented by `$2j$' points on a unit sphere~\cite{majorana-nc-1932}.

This concept has proved useful in various experimental contexts, such as for representing  states of polarization of light~\cite{hannay-jmo-1998}, for characterizing the symmetry of the order parameter in spinor Bose-Einstein condensates \cite{Suominen2007, Nakahara-njp-2018}, and for the decomposition of quantum gates used in nuclear magnetic resonance into experimentally implementable pulses ~\cite{dogra-jpb-2018}. In theoretical quantum information, the Majorana representation has enabled the geometric study of 
symmetric multi-qubit states~\cite{martin-pra-2010, devi-qip-2012}, the construction of geometrically mutually unbiased bases and symmetric informationally complete positive operator valued measures ~\cite{aravind-arxiv-2017}, the calculation  of the spectrum of the Lipkin-Meshkov-Glick model \cite{RibeiroVidal2007,RibeiroVidal2008}
the study of Berry phases \cite{Liu2016,Bruno2012}, and the  classification of high-dimensional entanglement~\cite{Liu2019, Ribeiro2011, devi-qip-2012}. It has also inspired the search for alternative geometric representations of quantum states \cite{arvind-jpa-1999,planat-ijtp-2008, makela-ps-2010, planat-geom-2011, ashourisheikhi-ijqi-2013, goyal-jpa-2016}. 

Our goal in this paper is to obtain and study the Majorana representation of single-qutrit dynamics for the stimulated Raman adiabatic passage (STIRAP) and for its superadiabatic version (saSTIRAP). STIRAP is a well-known protocol~\cite{bergmann-rev-1998,nikolay-arpc-2001}, widely used to perform
certain non-trivial quantum operations such as laser-induced population exchange between the energy levels of 
atoms and molecules. In circuit quantum electrodynamics, the protocol for STIRAP pulses
was first benchmarked in Ref.~\cite{kumar-nature-2016}, where it was experimentally realized 
using the first three energy levels of a transmon, and demonstrating population 
exchange between the energy levels using microwave fields. With the emergence of more and more controllable multilevel systems, and stimulated by research into quantum technologies, the applications of STIRAP are likely to expand \cite{Bergmann_2019}.  

However, adiabaticity
requires ideally infinitely long operation times, and therefore for realistic finite-time 
experimental conditions one expects a trade-off between time and fidelity. 
It is possible to accelerate STIRAP without loss of fidelity by the use of an additional drive (referred to as counterdiabatic) which exactly cancels the non-adiabatic excitations ~\cite{Fleischhauer99, berry-jpa-2009, Chen10, luigi-pra-2014}. This superadiabatic drive, also known as 
transitionless driving or assisted adiabatic passage has recently been experimentally implemented in
superconducting transmons~\cite{antti-science-2019}.
The superadiabatic STIRAP is a specific form of quantum control from the larger class of shortcuts to adiabaticity \cite{Torrontegui13}, which have found a wide variety of applications, for example  in electron transfer in quantum dots~\cite{nakahara-prb-2018},
state transfer in  nitrogen vacancy centers~\cite{jingfu-prl-2013},
 for the design of fast and error-resistant single-qubit ~\cite{antti-NOTgate-2018} and two-qubit \cite{YangYu2019} gates,  and for the quantum simulation of spin systems \cite{antti-arxiv-2017}.  Here we show that the dynamics induced by the counterdiabatic correction applied to the STIRAP protocol has a very intuitive picture when represented geometrically, as it brings the Majorana stars closer to the path of the dark state.

The paper is organized as follows. We begin with a brief introduction to the Majorana 
representation for a qutrit, see Sec.~\ref{intro}. Here we include the representation of the qutrit  by a symmetrized two-qubit state, and we define the spin-1 angular momentum vector. Sec.~\ref{sec:adia} describes STIRAP, 
followed by the dark-state dynamics on the Majorana sphere. We show here that the angular momentum vector provides a useful measure of nonadiabaticity, and we give also
a derivation of the superadiabatic protocol using the Majorana polynomial. Results from various simulations with experimentally
feasible parameters are presented in Section~\ref{sec:simulations}, including also the case of mixed states. We end with a review of the results and concluding remarks in Sec.~\ref{sec-conclusions}.

\section{Majorana representation}
\label{intro}

In 1932 Ettore Majorana introduced a representation of states with any angular momentum, as a starting point of his solution to the problem of atoms in a magnetic field \cite{majorana-nc-1932}. Consider the standard basis $\{|jm\rangle\}$ of angular momentum states, where $j$ and $m$ are the quantum numbers of the total angular momentum and its $z$-axis component. Specifically, setting the units to $\hbar = 1$ for convenience, we have $J_{z}|jm\rangle =  m|jm\rangle$ and $\mathbf{J}^2|jm\rangle = j(j+1)|jm\rangle$. The Majorana polynomial appears naturally when the spin-coherent representation of angular momentum states is considered. Writing $\mathbf{J} = J_{x}\hat{\mathbf{e}}_{x} +  J_{y}\hat{\mathbf{e}}_{y} + J_{z}\hat{\mathbf{e}}_{z}$, where $\hat{\mathbf{e}}_{x,y,z}$ are the unit vectors along the axes, we notice that the average of the angular momentum operator in the state $|jj \rangle$ equals the total angular momentum 
$\hbar j$, $\langle jj|\mathbf{J}| jj\rangle = j\hat{\mathbf{e}}_{z}$. We can obtain a vector with the same property
but oriented along an arbitrary direction $\hat{\mathbf{n}}$
with parametrization $\hat{\mathbf{n}} = (\cos\varphi\sin\theta , \sin\varphi\sin\theta , \cos\theta)$ in terms of spherical angles ($\theta, \varphi$), by an appropriate rotation of $|jj \rangle$, 

\begin{equation}
|j,\hat{\mathbf{n}}\rangle = e^{-i \theta \left(- \sin\varphi J_{x} + \cos\varphi J_{y}\right)}|jj\rangle .\label{spincoherent}
\end{equation} 
Indeed this transformation implements a counterclockwise rotation by an angle $\theta$ around the axis $(-\sin\varphi , \cos \varphi , 0)$, which brings $\hat{\mathbf{e}}_{z}$ along 
$\hat{\mathbf{n}}$. To obtain an expansion in the $|jm\rangle$ basis, we introduce the lowering and raising operators $J_{\pm} =J_{x} \pm i J_{y}$, and we rewrite Eq. (\ref{spincoherent}) accordingly,
\begin{equation}
|j,\hat{\mathbf{n}}\rangle = e^{\frac{1}{2}\theta \left( e^{i\varphi} J_{-} - e^{-i \varphi} J_{+}\right)}|jj\rangle. 
\end{equation} 
which resembles the standard form of Glauber bosonic coherent states. 
Next, we make use of the SU (2) algebra identity 
\begin{equation}
e^{\frac{1}{2}\theta \left( e^{i\varphi} J_{-} - e^{-i \varphi} J_{+}\right)}  = e^{\zeta J_{-}} e^{-\ln(1 + |\zeta |^2)J_{z}}e^{-\zeta^{*} J_{+}},
\end{equation}
where $\zeta = \tan \frac{\theta}{2}e^{i\varphi}$, 
to obtain $|j,\hat{\mathbf{n}}\rangle = (1 + |\zeta |^2)^{-j}e^{\zeta J_{-}}|jj\rangle $ 
Finally, by expanding  the exponential and employing the properties of the lowering operator we get
\begin{eqnarray}
|j,\hat{\mathbf{n}}\rangle &=&  \sum_{m=-j}^{j} \sqrt{\frac{(2j)!}{(j+m)!(j-m)!}}\left(\cos \frac{\theta}{2}\right)^{j+m}
\left(\sin \frac{\theta}{2}\right)^{j-m}  \nonumber \\
~ & & ~~~~~~~~~~~~~~~~~~~~~~~~~~~~~~~~~~\times e^{i(j-m)\varphi}|jm\rangle .
\end{eqnarray}
Now, given a general state
\begin{equation}
|\Psi \rangle = \sum_{m=-j}^{j} c_{m} |jm\rangle, \label{eq:stategeneral}
\end{equation}
where $c_m$ are complex amplitude probabilities normalized as $\sum_{m=-j}^{j}|c_{m}|^2 =1$,
the complex amplitude probability for transitions to the spin-coherent state oriented along $-\hat{\mathbf{n}}$ is obtained as 
\begin{equation}
\langle j, -\hat{\mathbf{n}}|\psi\rangle = \sqrt{(2j)!}\left(\cos \frac{\theta}{2}\right)^{2j} e^{-2ij\varphi}P_{|\psi\rangle}(\zeta ).
\end{equation}      
Here we define the Majorana polynomial using the same notations as in the original paper \cite{majorana-nc-1932},
\begin{equation}
P_{|\psi\rangle}(\zeta ) = \sum_{r=0}^{2j} a_{r}\zeta^{2j -r},
\end{equation}
in the complex variable $\zeta = \tan \frac{\theta}{2}e^{i \varphi}$.  The relation between the coefficients of the polynomial and the complex amplitudes of the state is
\begin{equation}
a_{r}= \frac{(-1)^r}{\sqrt{(2j-r)! r!}}c_{j-r}. \label{coeff}
\end{equation} 
Applying the fundamental theorem of algebra, it follows that $P_{|\psi\rangle}(\zeta )$ has $2j$ roots, which correspond to points in the complex $xOy$ plane.  

Next, we notice that if we represent these roots by the angles $\theta$ and $\varphi$ as above, they can be represented by an inverse stereographic projection on the Riemann sphere, with respect to the South Pole as the reference. It is straightforward to prove geometrically that a line that connects the South Pole with a point  $\zeta = \tan \frac{\theta}{2}e^{i \varphi}$ in the $xOy$ plane will intersect the unit sphere at a point with spherical coordinates $(\theta,\varphi)$. This unique
configuration of $2j$ Majorana stars is invariant under rotations and achieves 
a geometrical representation of $\vert{\Psi} \rangle$ called Majorana constellation.
For instance,
$(\theta= \pi$, $\varphi=0)$ is a Majorana star on the South Pole and a point at infinity in the complex plane, 
while $(\theta=0$, $\varphi=0)$ is a Majorana star placed at the North Pole and corresponds to the center of the complex plane. The Majorana representation of the state $|jm\rangle$ consists of $j+m$ stars at the North Pole and $j-m$ stars at the South Pole. Also, for a general state of a spin-1/2, $\cos (\theta /2)|1/2,1/2\rangle + \sin (\theta /2)e^{i\varphi}|1/2,-1/2\rangle$ one could verify immediately that we recover the standard Bloch representation, namely that the Majorana star is a point on the Bloch sphere, with spherical coordinates $(\theta, \varphi)$.

Operators can also be represented in the Majorana picture by the use of differentials in the complex variable $\zeta$.
For example, let us consider the $x,y,z$-components $J_x, J_y, J_z$ of the 
angular momentum operator $\mathbf{J} = J_{x}\hat{\mathbf{e}}_{x} + 
J_{y}\hat{\mathbf{e}}_{y} + J_{z}\hat{\mathbf{e}}_{z}$
given by
$J_x=(|0\rangle \langle 1| + |1\rangle \langle 2| + \mathrm{h.c.}
)/\sqrt{2}$, $J_y=(-i|0\rangle \langle 1| - i|1\rangle \langle 2| 
+  \mathrm{h.c.}))/\sqrt{2}$, and 
$J_z=(|0\rangle \langle 0| - |2\rangle \langle 2|)/2$. 
We then get by using the properties of spin-coherent states \cite{PhysRevA.40.6800} and the 
definitions above
\begin{eqnarray}
J_{x}(\zeta ) &=& \frac{1}{2} (-2\zeta + \zeta^{2}\partial_{\zeta} - \partial_{\zeta}), \label{eq:Jx}\\
J_{y} (\zeta ) &=& \frac{1}{2i}( - 2 \zeta + \zeta^{2}\partial_{\zeta} + \partial_{\zeta} ), \label{eq:Jy}\\
J_{z}(\zeta ) & =& - 1 + \zeta\partial_{\zeta}. \label{eq:Jz}
\end{eqnarray}
These can be verified by taking a general state $|\psi\rangle$ as in Eq. (\ref{eq:stategeneral}), constructing the Majorana polynomial $P_{|\psi \rangle}(\zeta )$, acting with Eqs. (\ref{eq:Jx},\ref{eq:Jy},\ref{eq:Jz}), and showing that the polynomials thus generated
$J_{x,y.z}(\zeta ) P_{|\psi \rangle}(\zeta )$ are the Majorana polynomials of  $J_{x,y.z}|\psi \rangle$.

Now we can introduce the geometrical picture of a qutrit ($j=1$), which is a three level quantum 
system -- thus its Majorana geometrical representation consists of two Majorana stars. 
An arbitrary state of a qutrit in the computational basis $\{ |0\rangle, |1\rangle , |2\rangle \}$
is given by
\begin{equation}
 \vert{\Psi} \rangle = \mathcal{C}_{0} \vert{0} \rangle + \mathcal{C}_{1}
\vert{1} \rangle + \mathcal{C}_{2} \vert{2} \rangle, \label{eq:qutrit_state}
\end{equation}
where 
$\mathcal{C}_{0}$, $\mathcal{C}_{1}$, and $\mathcal{C}_{2}$ are complex probability amplitudes.

The second-degree Majorana polynomial $P_{|\phi\rangle} (\zeta ) = a_0 \zeta^2 + a_1 \zeta + a_2$ 
associated with the state of a qutrit is obtained by identifying the $j=1$ basis $|jm\rangle$ ($m=-1, 0, 1$) with the computational basis, $|1m\rangle = |1-m\rangle$ ({\it i.e.} $\mathcal{C}_0 = c_{1}$,  $\mathcal{C}_1 = c_{0}$, $\mathcal{C}_2 = c_{-1}$), and the coefficients are given from Eq. (\ref{coeff})
as $a_2=\mathcal{C}_2/\sqrt{2}$, $a_1=-\mathcal{C}_1$, and $a_0=\mathcal{C}_0/\sqrt{2}$. The roots $\zeta_{k} = \tan \frac{\theta_{k}}{2}e^{i\varphi_{k}}$, $k\in\{1,2\}$ can be found by solving a second-order equation, and the polynomial takes the form $P_{|\phi\rangle} (\zeta ) = a_{0}(\zeta - \zeta_{1})(\zeta - \zeta_{2})$.
For the state $|0\rangle$ both Majorana points lie on the North Pole,  
 $|2\rangle$ has both points lying on the South Pole,
 while $|1\rangle$ is represented by one point on the North Pole and another point on the South Pole.  We define the distance between two Majorana points as $\eta=\cos^{-1}(\hat{\mathbf{S}}_1\cdot\hat{\mathbf{S}}_2)=\cos^{-1}[\sin\theta_1 \sin\theta_2 \cos(\varphi_1-\varphi_2)+\cos\theta_1\cos\theta_2]$, where the Majorana stars $\hat{\mathbf{S}}_k$ have respective spherical coordinates $(\theta_{k}, \varphi_{k})$, $k\in \{1,2\}$. This distance can be interpreted as a direct measure of entanglement between two qubits in the symmetrized-state representation of the qutrit \cite{Ribeiro2011,Liu2019}. Explicitly, any spin-1 state  -- and, by the identification above, any qutrit state Eq. (\ref{eq:qutrit_state}) -- can be represented as a symmetric combination of two spin-1/2 (qubit) states
$|\psi_{k}\rangle =\cos (\theta_{k}/2) |1/2,1/2\rangle + \sin (\theta_{k}/2) e^{i\varphi_{k}}|1/2,-1/2\rangle$ (with $k=1,2$), as 
\begin{equation}
|\Psi \rangle = \frac{1}{\sqrt{2[1+ \cos^{2}(\eta/2)]}}\sum_{\sigma} |\psi_{\sigma (1)}\rangle \otimes |\psi_{\sigma (2)}\rangle \label{eq:repr}
\end{equation}
where $\sigma$ are permutations of the indices $k$. Using the definitions above, the Majorana stars of this state coincides with the two points with spherical coordinates $(\theta_{k}, \varphi_{k})$, representing the two qubits in the standard Bloch representation.

It follows immediately that the concurrence between the two qubits is measured by the angle between the Majorana stars,
\begin{equation}
\mathrm{C_{|\Psi\rangle}} = \frac{\sin^2(\eta /2)}{1+  \cos^{2} (\eta /2)}, \label{eq:concurrence}
\end{equation}
which is obtained by applying Wooters' formula \cite{Horodecki2009}.

Majorana representation is complete and unique up to a global phase.
Given a qutrit state $\vert \Psi \rangle$, one can always find a pair of points 
representing the state on the Majorana sphere. Alternatively, corresponding to any arbitrary 
pair of points on the Majorana sphere, one can construct a unique $\vert \Psi \rangle$ in 
the three-dimensional Hilbert space of a qutrit.

We now show that there is a deeper connection between the average of the angular momentum and the entanglement properties in the symmetrized state representation of spin 1. Starting from Eq. (\ref{eq:repr}), we take the partial trace of the total density matrix $|\Psi\rangle\langle \Psi |$ over one of the spin-1/2. The resulting reduced density matrix $\rho$ does not depend on which spin-1/2 we traced out, and it reads
\begin{equation}
\rho = \frac{\mathbb{I}_2}{2} + \frac{1}{2} \mathbf{r}\boldsymbol{\sigma},
\end{equation}
where
\begin{equation}
\mathbf{r} = \frac{1}{1 + \cos^2 (\eta /2)}(\hat{\mathbf{S}}_{1} + \hat{\mathbf{S}}_{2}). \label{eq:rvector}
\end{equation}
We see that the vector $\mathbf{r}$ that parametrizes the reduced density matrix can be obtained, up to a normalization factor,  by the vectorial addition of the Majorana vectors. 
We can now calculate immediately the average 
\begin{equation}
\langle \mathbf{J} \rangle = \langle J_x \rangle \hat{\mathbf{e}}_{x} + \langle J_y \rangle \hat{\mathbf{e}}_{y} + \langle J_z \rangle \hat{\mathbf{e}}_{z} \label{eq:mag_sigma}
\end{equation}
of the total angular momentum on the state $\vert \Psi \rangle$, where $\langle J_k \rangle=\langle \Psi \vert J_k \vert \Psi \rangle$, $k \in \{ x, y, z \}$.
Since this is a sum of the angular momenta $\boldsymbol{\sigma}/2$ of each of the two qubits, we get
\begin{equation}
\langle \mathbf{J} \rangle = \mathrm{Tr} (\rho \boldsymbol{\sigma}) = \mathbf{r}. \label{eq:aver}
\end{equation}
Thus, the average vector angular momentum is given by the vector $\mathbf{r}$ that parametrizes the reduced density matrix. 
This leads to a geometric representation of the average angular momentum as the vector
\begin{equation}
\langle \mathbf{J} \rangle = \frac{2}{3 + \mathbf{S}_{1}\cdot\mathbf{S}_{2}}\left(\mathbf{S}_{1} + \mathbf{S}_{2}\right), \label{eq:geo}
\end{equation}
with length
\begin{equation}
|\langle \mathbf{J}\rangle |= \frac{2\vert\cos (\eta/2)\vert}{1+\cos^2(\eta/2)}, \label{eq:mag}
\end{equation}
This vector is manifestly 
invariant under rotations generated by $J_x, J_y, J_z$ ~\cite{dogra-jpb-2018}. It is interesting to point out that $\mathbf{S}_{1}$ and 
$\mathbf{S}_{2}$ can themselves be regarded as vectors resulting from averaging the Pauli matrices over the constituent spin-1/2 states, $\mathbf{S}_{k}=\langle \psi_{k}| \boldsymbol{\sigma} |\psi_{k}\rangle$, $k\in \{1,2\}$. Thus, Majorana representation offers a remarkable way of writing the quantum-mechanical addition of angular momenta as a geometrical vector addition.

Finally, the purity of the reduced state $1-Tr(\rho^2) = 1/2 - |\mathbf{r}|^2/2$ can be used to calculate the concurrence $C_{|\Psi \rangle}=\sqrt{2[1-Tr(\rho^2)]}$, with the result 
$C_{|\Psi \rangle} = \sin^2 (\eta/2)/[1 + \cos^2 (\eta/2)]$, identical with the one calculated above in Eq. (\ref{eq:concurrence}).

Next, we closely follow the dynamics of the Majorana stars and its corresponding spin-1 average vector, as our single qutrit pure 
state undergoes unitary evolution generated by STIRAP and superadiabatic(sa)-STIRAP 
Hamiltonians.

%%%%%%%%%%%%%%%%%%%%%%%%%%%%%%%%%%%%%

\section{Adiabatic and superadiabatic processes for qutrits}
\label{sec:adia}

\subsection{STIRAP \label{sec-STIRAP}} 
The Stimulated Raman Adiabatic Passage (STIRAP) 
is a fundamental quantum mechanical process for transferring 
population between the single-qutrit ground state $|0\rangle$ and the second excited state 
$|2\rangle$ without populating the intermediate state $|1\rangle$.
This is achieved via a counter-intuitive sequence~\cite{bergmann-rev-1998,nikolay-arpc-2001}, whereby
the driving between levels $|1\rangle-|2\rangle$ commences prior to that of
$|0\rangle-|1\rangle$. Artifical atoms based on Josephson junctions are suitable for this task, since they are essentially  multilevel systems where the transitions can be driven by microwaves \cite{Wei2008,pino_PRB2009,Paraoanu2014,Buluta_2011,Nori_review_2011}. For example, consider a transmon irradiated with two such GHz-frequency fields with time-dependent Rabi couplings
$\Omega_{01}(t)$ and $\Omega_{12}(t)$, phases $\phi_{01}$ and $\phi_{12}$, and driving frequencies $\omega_{01}^{(\Omega)}$ 
and $\omega_{12}^{(\Omega)}$,  coupling the levels $|0\rangle-|1\rangle$
and $|1\rangle-|2\rangle$ respectively \cite{Falci2017}. These microwave fields can be
slightly detuned from the respective transmon transition frequencies 
$\omega_{01}$ and $\omega_{12}$, and the robustness of STIRAP can be  studied with respect to these detunings \cite{kumar-nature-2016}. However, for simplicity we consider here the fully resonant case. 
Then, under the rotating wave approximation, the effective Hamiltonian governing STIRAP 
is given by
\begin{eqnarray}
\rm{H}_0 &=& \frac{\hbar}{2} \Omega_{01} e^{i\phi_{01}}|0\rangle\langle 1| + \frac{\hbar}{2}  \Omega_{12}e^{i\phi_{12}}|1\rangle\langle 2| + \mathrm{h.c.} 
\end{eqnarray}

The time-varying amplitudes of the microwave drives are chosen of Gaussian form in time-domain,  
with the same standard deviation $\sigma$ -- and therefore the same 
width.  They are separated in time by an amount $t_s$, which, when negative, leads to the counterintuitive sequence. The corresponding Rabi couplings read
\[ \Omega_{01}(t) = \bar{\Omega}_{01} e^{-t^2/2\sigma^2} \quad \textrm{and} \quad  \Omega_{12}(t) = \bar{\Omega}_{12} e^{-(t-t_s)^2/2\sigma^2}.\] 
An adiabatic evolution is ideally infinitely slow and would require the system to be in 
an eigenstate of the instantaneous Hamiltonian at all times. By fixing  the gauge to $\phi_{01}=\phi_{12}=0$ \cite{antti-arxiv-2017} we obtain the eigenvectors of $\rm{H}_0$ in the form
 \begin{eqnarray}
  \vert n_{\pm}\rangle &=& \frac{1}{\sqrt{2}}\left[\sin \Theta \vert 0 \rangle + \cos \Theta \vert 2 \rangle \right] \pm \frac{1}{\sqrt{2}}\vert 1 \rangle,  \nonumber \\
  \vert D \rangle &=& \cos \Theta \vert 0 \rangle - \sin \Theta \vert 2 \rangle, \label{eq:eigenv}
\end{eqnarray}
with eigenvalues $E_{\pm}=\pm\hbar\sqrt{\Omega_{01}^2  + \Omega_{12}^2 }/2$
and $ E_{D}=0$ respectively. 
Here the STIRAP angle $\Theta$ is given by
\begin{equation}
 \tan \Theta = \frac{\Omega_{01}}{\Omega_{12}}. \label{eq:theta}
\end{equation}
A convenient choice for the state to be followed adiabatically is the dark state $\vert D \rangle$, which does not contain any component from
the intermediate level $\vert 1 \rangle$.
Interestingly, $\vert D \rangle$ is the same as the single-qutrit canonical state \cite{dogra-jpb-2018}, up to a change of variables $\Theta \rightarrow \Theta - \pi /2$. The canonical state is a single-parameter state which spans the entire qutrit Hilbert space under SO(3) rotations generated by
$J_x$, $J_y$, and $J_z$. These unitary operations rotate the 
Majorana stars like a rigid body in three-dimensional real space($\mathbb{R}^3$)~\cite{dogra-jpb-2018}; thus, any qutrit state can be parametrized with one parameter from the canonical state and three parameters from the three rotations. In other words, any arbitrary state of a qutrit can be obtained from the set of canonical states (or dark states) via SO(3) rotations.

\begin{figure}
 \centering
 \includegraphics[scale=1,keepaspectratio=true]{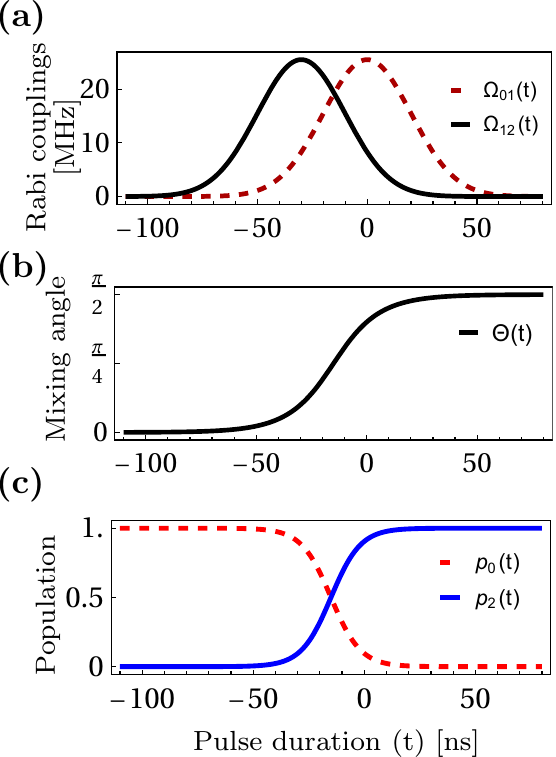}
 % pulse-fig.pdf: 0x0 pixel, 300dpi, 0.00x0.00 cm, bb=
 \caption{(a) Variation of the Rabi couplings  with time, (b) corresponding
 variation of $\Theta$ and (c) the probabilities $p_0=\cos^2 \Theta$ and $p_2=\sin^2 \Theta$, plotted 
 in real time.}
 \label{fig-timevariation}
\end{figure}

 Under adiabatic evolution, as a consequence of the adiabatic theorem \cite{Born28}, the system remains in the state 
$\vert D \rangle$ at all times. To
transfer the population from $\vert 0 \rangle$ to $\vert 2 \rangle$, one can initialize the 
system in a state corresponding to $\Theta=0$ at $t=t_i$, which eventually transforms into $\Theta=\pi/2$ at $t=t_f$,
with the total duration of the sequence being $\rm{T}=t_f-t_i$, with $t_i=-110$ ns and $t_f=80$ ns as shown in Fig.~\ref{fig-timevariation}. 
The Gaussian pulse profiles $\Omega_{01}(t)$ and $\Omega_{12}(t)$ are
shown in Fig.~\ref{fig-timevariation}(a) for $t_s = -30$ ns, $\sigma = 20$ ns, and $\bar{\Omega}_{01}=\bar{\Omega}_{12}= 2\pi \times 25.5$ MHz.
The corresponding variation of $\Theta$ from $0$ to $\pi/2$ is clearly reflected from Fig.~\ref{fig-timevariation}(b), and 
the probabilities of occupation of the energy levels in terms of the squares of the absolute 
values of the coefficients, $p_0=\cos^2 \Theta$ and $p_2=\sin^2 \Theta$ are plotted in Fig.~\ref{fig-timevariation}(c).
The rate of change of the populations is
\begin{equation}
  \dot{p_0}= - \dot{p_2}= -\dot{\Theta} \sin 2\Theta , \label{popRate}
\end{equation}
which tends to zero as the mixing angle approaches its extreme values $\Theta=0$ and $\pi/2$. The rate of change of populations 
attains its maximum value at $t=\rm{T}/2 = (t_f-t_i)/2$. 
A geometrical picture of this dynamics may be obtained on the Majorana sphere, which is discussed further in Sec.~\ref{sec-MajD}.
% %%%%%%%%%%%%%%%%%%%%%%%%%%%%%%%%%%%%%%%%%%%%%%%%%%
\subsection{Majorana representation of the dark state $\vert D \rangle$ \label{sec-MajD}}
Recalling that the dark state is $\vert D \rangle=\cos \Theta \vert 0 \rangle - \sin \Theta \vert 2 \rangle$, 
the Majorana polynomial is $P_{|D\rangle }(\zeta) = (1/\sqrt{2})(\cos \Theta \zeta^2 - \sin \Theta)$. The Majorana representation of $\vert D \rangle$ consists of two stars $S_1(\theta,\pi)$ and $S_2(\theta,0)$,
where $\theta=\pi-2\tan^{-1}\sqrt{\cot \Theta}$. These points lie in the $xz$-plane with Cartesian co-ordinates $(\pm x_D, y_D, z_D)$, such that
\begin{eqnarray}
 x_D &=&\frac{\sqrt{2\sin{2 \Theta}}}{\cos{\Theta}+\sin{\Theta}}, \nonumber \\
 y_D &=& 0, \nonumber \\
 z_D &=& \frac{\cos{\Theta}-\sin{\Theta}}{\cos{\Theta}+\sin{\Theta}},
\end{eqnarray}
where $\Theta \in [0, \pi/2]$. 

%%%%%%%%%%%%%%%%%%%%%%%%%%%%%%%%%%%%%%%%%%%%%%
\begin{figure}[h!]
\centering
\includegraphics[scale=0.9]{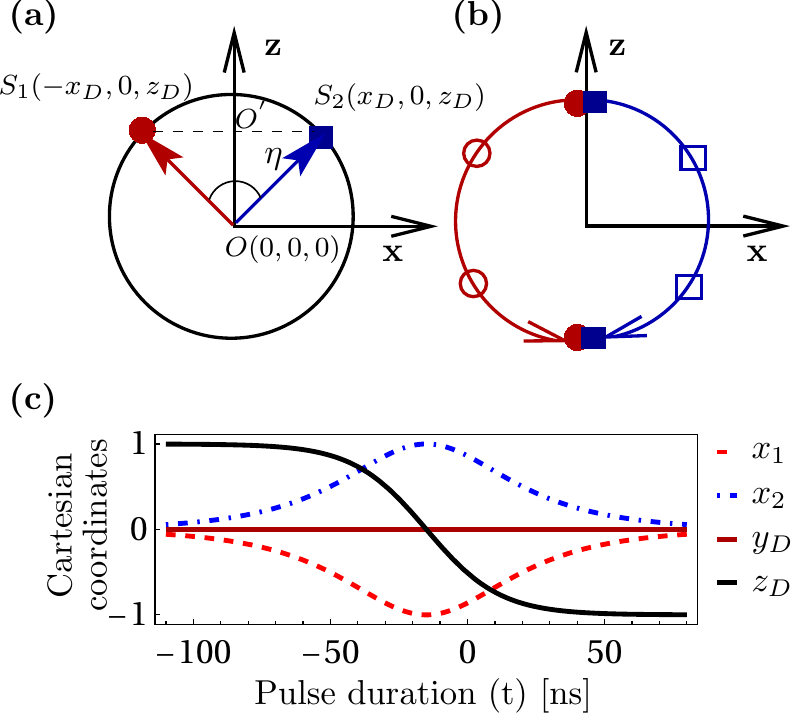}
\caption{A cross-section of the Majorana sphere in the plane $y=0$, which shows 
 the Majorana representation of the dark state $\vert D \rangle$ and its dynamics under 
 STIRAP Hamiltonian. (a) Majorana stars are labeled by
 $S_1(x_1, y_1, z_1)$ and $S_2(x_2, y_2, z_2)$ with red circle and blue square respectively, such that
 $-x_1=x_2=x_D$, $y_1=y_2=0$, and $z_1=z_2=z_D$. (b) Trajectory of both the points on the Majorana sphere as $\Theta$ varies from $0$ to $\pi/2$.
 (c) Variation in real time of the co-ordinates of Majorana points 
$S_1(x_1,y_D,z_D)$ and $S_2(x_2,y_D,z_D)$ is plotted as $\Theta$
changes from $0$ to $\pi/2$ as in Fig.~\ref{fig-timevariation}(b).\label{Fig-points} \label{canonical-rep}}
\end{figure}
Majorana stars representing $\vert D \rangle$ lie on the great circle in the plane $y=0$, with the same latitude $\theta$.
The angle $\eta$ formed by the vectors $\mathbf{S}_1$ and $\mathbf{S}_2$ 
is given by  $\eta=2\theta$.
As $\Theta$ varies from
$0 \leq \Theta \leq \pi/2$, $\eta$ assumes values from
$0$ to $\pi$ and then $0$ again. 
Thus the two points $S_1$ and 
$S_{2}$ lie all along the great circle in the plane $y=0$
as shown in
Fig.~\ref{canonical-rep}(a). This readily leads to the fact that the 
dynamics of $\vert D \rangle$ under the STIRAP Hamiltonian is confined to 
the plane $y=0$ on the Majorana sphere {\it i.e.} the longitudes remain unchanged while sweeping the latitudes 
$\theta_1=\theta_2=\theta$ from $0$ to $\pi$, see Fig.~\ref{canonical-rep}(b). Under STIRAP, both stars
representing $\vert D \rangle$ on the Majorana sphere move symmetrically  
with varying $\Theta$, and any deviation from this behavior signals a loss of adiabaticity. The trajectories of these Majorana
stars in the co-ordinate space under the action of STIRAP are shown in Fig.~\ref{Fig-points}(c).
The trajectories are plotted in time with the same scales as described in Sec.~\ref{sec-STIRAP}.

\par
The dark state can also be represented using symmetrized
spin-1/2 states, as described in Section II, in the form $|D\rangle = \cos \Theta |1/2, 1/2 \rangle |1/2, 1/2 \rangle - \sin \Theta |1/2, -1/2 \rangle |1/2, -1/2 \rangle$. Thus, the resulting symmetrized dark state is manifestly in the Schmidt form \cite{Horodecki2009}. The concurrence is obtained as $\mathcal{C}_{|D\rangle} = \sin 2\Theta$ and the $r$-parameter of the reduced density matrix is $r = \cos 2 \Theta$. Given a qutrit state $|\psi\rangle$  with Majorana polynomial roots $\zeta_{k} = \tan (\theta_{k}/2) \exp( i \varphi_{k})$, where $k\in \{ 1,2 \}$, the fidelity of this state with respect to the dark state is obtained as
\begin{eqnarray}
F &=& |\langle \psi | D\rangle |^2 \label{eq:fidelity} \\
  &=& \frac{1}{2[ 1+ \cos^2(\eta / 2)]}}{[
  1 + (\cos\theta_{1} + \cos\theta_{2}) \cos 2\Theta +  \nonumber \\
 & & + \cos\theta_{1} \cos\theta_{2} - \sin\theta_{1}\sin\theta_{2}\sin 2\Theta \cos (\varphi_1 + \varphi_2 )]. \nonumber
  \end{eqnarray}

The Majorana representation of the spin-$1$ average vector in the dark state $\vert D \rangle$ is obtained using
the bisector of angle $\eta$ as shown with $\mathbf{OO'}$ in Fig.~\ref{canonical-rep}(a), such that  
\begin{equation}
 \langle \mathbf{J}\rangle =\frac{2\mathbf{OO'}}{1+|\mathbf{OO'}|^2}. \label{eq:J1}
\end{equation}
We note that this relation is in fact valid for any general qutrit state, since it follows from Eq.~(\ref{eq:geo}) with $\mathbf{OO'}=(\hat{\mathbf{S}}_{1} + \hat{\mathbf{S}}_{2})/2$.
Specifically for the dark state $\vert D \rangle$, we get from 
Eq~(\ref{eq:mag_sigma}) that $\langle J_x \rangle=0$, $\langle J_y \rangle=0$,
and $\langle J_z \rangle=\cos 2\Theta$.

The magnitude of the spin-$1$ average vector is given by $\vert\langle \mathbf{J}\rangle \vert=\vert\cos 2{\Theta}\vert$, which varies with time as
\begin{equation}
 \partial_{t}\vert\langle \mathbf{J}\rangle \vert =2\dot{\Theta} \sin2\Theta.
\end{equation}
This is twice the rate of change of populations $2\dot{p}_{2}(t)$ from Eq.(\ref{popRate}),
which puts in evidence that the Majorana geometrical picture 
reflects the rate of evolution of a quantum state with time.  Indeed, the co-ordinates of the Majorana stars vary slowly with time near the 
starting and end points, and change significantly faster in between, as 
expected for an adiabatic evolution.

The Majorana representation also offers a geometric tool for quantifying the degree of nonadiabaticity of STIRAP. Indeed, 
for a general state Eq.(\ref{eq:qutrit_state}) we have $\mathcal{C}_{1} =  \mathcal{C}_{0}(\zeta_{1} + \zeta_{2})/\sqrt{2}$, and therefore the non-adiabatic population on the intermediate state is $p_{1}=|\mathcal{C}_{0}|^2|(\zeta_{1} + \zeta_{2})|^{2}/2$. Thus, deviations from adiabaticity in the case of STIRAP are signaled by nonzero $x,y$-components of the vector $\hat{\mathbf{S}}_{1} + \hat{\mathbf{S}}_{2}$ or, equivalently, $\langle J_{x}\rangle$ and $\langle J_{y}\rangle$, see Eq. (\ref{eq:geo}). These deviations from zero can be immediately seen in the plots, as we will discuss extensively later. Thus, once the roots are calculated, this is a simple and exact method, in contrast to other approximate ways of quantifying the degree of nonadiabaticity based for example on the Dykhne-Davis-Pechukas technique \cite{PhysRevA.80.013417}. The well-established STIRAP local adiabaticity condition $\dot\Theta \ll \sqrt{\Omega_{01}^{2} + 
\Omega_{12}^{2}}$ \cite{bergmann-rev-1998} can be derived as well in the Majorana picture by noticing that 
$P_{|n_{\pm} \rangle}(\zeta ) = \frac{1}{2}\sin \Theta \zeta^2 \mp \frac{1}{\sqrt{2}}\zeta + \frac{1}{2}\cos\Theta$ and $\dot{P}_{|n_{\pm} \rangle}(\zeta ) = \frac{1}{\sqrt{2}}\dot{\Theta}P_{|D\rangle}(\zeta )$. Thus the nonadiabatic coupling is $|\langle n_{\pm}|\partial_{t}|D \rangle |= |\langle \dot{n}_{\pm}| D \rangle | = \dot{\Theta}/\sqrt{2}$, which has to be much smaller than the splitting between the dark state and the states $|n_{\pm}\rangle$, $|E_{\pm}|/\hbar  = \sqrt{\Omega_{01}^{2} + \Omega_{12}^{2}}$.

%%%%%%%%%%%%%%%%%%%%%%%%%%%%%%%%%%%%%%%%%%%%%%%%%
\subsection{Superadiabatic(sa)-STIRAP \label{sec-saSTIRAP}}
%%%%%%%%% Figure 3
\begin{figure}
 \centering
 \includegraphics[scale=1,keepaspectratio=true]{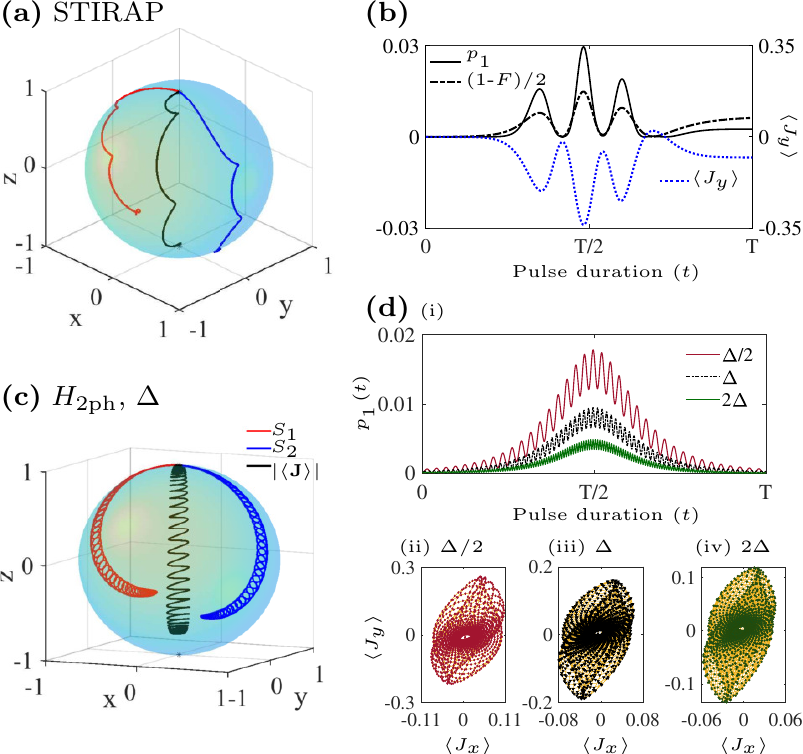}
 % Maj-saSTIRAP.pdf: 0x0 pixel, 300dpi, 0.00x0.00 cm, bb=
 \caption{
 The qutrit, initialized in the state $|0\rangle$ at the North Pole of the Majorana sphere, evolves under the STIRAP Hamiltonian and the two-photon drive $H_{2\rm{ph}}$, such that its
 	expected final state is -$|2\rangle$ -- the South Pole of the Majorana sphere.
 	(a) STIRAP - driven trajectories of the Majorana stars (in red and blue colors) and the spin-$1$ angular momentum vector (in black), and correspondingly (b) the population of the first excited state $p_1$ with solid black line, fidelity loss $(1-F)/2 \approx|\langle \psi(t)|n\pm \rangle|^2$ with dot-dashed black line, and the $y-$component of the angular momentum vector $\langle J_y \rangle$ with dotted blue line. Here $p_1$ and $(1-F)/2$ follow the left vertical scale, while $\langle J_y \rangle$ follows the right vertical scale. (c) Majorana trajectories resulting from the $H_{2\rm{ph}}$ drive with a detuning $\Delta=2 \pi \times 225$ MHz w.r.t. the single-photon drives. The plots in (d)(i) present the populations in state $\vert 1 \rangle$ resulting from  different detunings $\Delta/2$ (thick red line), $\Delta$ (dashed black line), and $2\Delta$ (thin green line) respectively under the $H_{2\rm{ph}}$ drive. The corresponding projections of the angular momentum vector $\langle \mathbf{\rm J} \rangle $ in $xy$-plane are shown in (d)(ii) with red diamonds, (iii) with black triangles, and (iv) with green circles respectively.
 \label{Fig:Maj-STIRAP-H2ph}}
\end{figure}
%%%%%%%%%%%%%%
It is interesting to observe on the Majorana sphere the dynamics of the 
quantum state under the combined effect of STIRAP Hamiltonian 
and a counter-diabatic drive. When acting together, this realizes the so-called superadiabatic(sa)-STIRAP.
The counderdiabatic part achieves the cancellation of spurious non-adiabatic excitations, leading to the expected final state with high 
precision at finite times. The counterdiabatic term in case of a 
three-level quantum system requires a very simple form ~\cite{antti-arxiv-2017},
\begin{equation} \label{eq:CD}
\rm{H}_{\mathrm{cd}} =\frac{\hbar}{2}\Omega_{02}e^{i \phi_{02}}
|0\rangle \langle 2 | + \mathrm{h.c.} 
\end{equation}
with $\Omega_{02}(t)=2\dot{\Theta}(t)$ and $\phi_{02}=\pi/2$, leading overall to a purely imaginary Rabi coupling. 
This is experimentally 
realized by a two-photon pulse resonant with the 0 - 2 energy level difference ~\cite{antti-arxiv-2017}, which simultaneously drives the $|0\rangle-|1\rangle$ and $|1\rangle-|2\rangle$ transitions
with detunings $\mp \Delta$ respectively, where $\Delta=(\omega_1-\omega_2)/2$. The Hamiltonian in the doubly-rotating frame 
driven by tones with frequencies $\omega_1$ and $\omega_2$ is given by
\begin{equation} \label{eq:h2ph_lambda}
\rm{H}_{2\rm{ph}} = \frac{\hbar}{2}\Omega_{2\rm{ph}}e^{(i \phi_{2\rm{ph}}-\Delta t)}|0\rangle \langle 1 | + \frac{\hbar\sqrt{2}}{2}\Omega_{2\rm{ph}}e^{(i \phi_{2\rm{ph}}+\Delta t)}|1\rangle \langle 2 | 
+ \mathrm{h.c.}
\end{equation}
with effective coupling $|\Omega_{2\rm{ph}}|=\sqrt{\sqrt{2}\Delta\Omega_{02}}$ and phase $\phi_{2\rm{ph}}=(\phi_{02}-\pi)/2$.

%%%%%%%%%%%%%

\subsection{Derivation of the superadiabaticity condition in the Majorana representation}

Before proceeding to investigate in detail the motion of Majorana stars under the saSTIRAP drive corresponding to realistic experimental conditions, we notice  that the Majorana representation enables an alternative derivation of the counteradiabatic condition. For simplicity, in  this subsection  we consider as before the all-resonant STIRAP case with the gauge \cite{antti-arxiv-2017} fixed to $\phi_{01}=\phi_{12}=0$, and we set $\phi_{02} = \pi /2$ by adjusting the phase of the two-photon pulse to $\phi_{2\rm{ph}}=- \pi/4$.
Thus, the available Hamiltonians are $\rm{H}_{0} =  \frac{\hbar}{2}\left[\Omega_{01} |0\rangle \langle 1| + \Omega_{12} |1\rangle \langle 2|\right] + \mathrm{h.c.}$ and $\mathrm{H}_{\mathrm{cd}} = \frac{i\hbar}{2} \Omega_{02}|0\rangle\langle 2| + \mathrm{h.c.}$ is generated as well \cite{antti-science-2019,antti-arxiv-2017}. These Hamitonians have a convenient Majorana representation in a frame defined by the unitary transformation which in the basis $\{|0\rangle$, $|1\rangle$, $|2\rangle \}$ has the form
\begin{equation}
R = \frac{1}{\sqrt{2}}
\left(\begin{array}{lll}
1 & 0 & 1 \\ 0 & \sqrt{2} & 0 \\  i & 0 & -i
\end{array}\right)
\end{equation}
with $R^{\dag}R = R R^{\dag} = \mathbb{I}$. In the $R$ - frame, the wavefunctions become $|\psi\rangle^{(R)} = R^{\dag}|\psi\rangle$, while  the 
Hamiltonians transform as $\mathrm{H}_{0}^{(R)} = R^{\dag} \mathrm{H}_{0} R$ 
%= \frac{\hbar}{2}\left[\Omega_{01}(t) \Lambda_{1}^{(R)} +\Omega_{12}(t) \Lambda_{6}^{(R)}\right]$ 
and  $\mathrm{H}_{\mathrm{cd}} = R^{\dag} \mathrm{H}_{\mathrm{cd}} R$. 
%= -\frac{\hbar}{2} \Omega_{02}(t)\Lambda_{5}^{(R)}$. 
It is straightforward now to verify that $J_{x} = R^{\dag}(|0\rangle \langle 1| + |1\rangle \langle 0|) R $, 
$J_{y} = R^{\dag}(|1\rangle \langle 2| + |2\rangle \langle 1|)R $, and $J_{z} = R^{\dag} (-i|0\rangle \langle 2| + i|2\rangle \langle 0|)R $ are respectively the $x$, $y$, and $z$ angular momenta $J_{x}, J_{y}, J_{z}$ in the standard representation (see Section II).
Thus, in the $R$-frame the Hamiltonian $\mathrm{H}_{0}^{(R)} + \mathrm{H}^{(R)}_{\mathrm{cd}}$ with 
\begin{equation}
\mathrm{H}_{0}^{(R)} = \frac{\hbar \sqrt{\Omega_{01}^{2} + \Omega_{12}^{2}}}{2}\left( \sin\Theta J_{x} + \cos \Theta J_{y}\right)
\end{equation}
and
\begin{equation}
\mathrm{H}_{\mathrm{cd}}^{(R)} = -\frac{\hbar \Omega_{02}}{2}J_{z}
\end{equation}
describes a spin-1 particle in a vector magnetic field 
$(\Omega_{01}(t), \Omega_{12}(t), -\Omega_{02}(t))$. 
Here we recall that $\tan \Theta  = \Omega_{01} /\Omega_{12}$.

Consider now a general state $|\psi\rangle^{(R)}$, to which we can associate a Majorana polynomial $P_{|\psi\rangle^{(R)}}$. Using the Majorana representation for the angular momentum Eqs. 
(\ref{eq:Jx},\ref{eq:Jy},\ref{eq:Jz})
we get
\begin{equation}
H_{0}^{(R)}(\zeta ) = -\frac{i\hbar\sqrt{\Omega_{01}^{2} + \Omega_{12}^{2}}}{4} \left[e^{i\Theta}(-2 \zeta + \zeta^2\partial_{\zeta}) + e^{-i \Theta}\partial_{\zeta}\right],
\end{equation} 
and 
\begin{equation}
H_{\textrm{cd}}^{(R)} (\zeta )= \frac{\hbar \Omega_{02}}{2} (\zeta \partial_{\zeta} - 1 ).
\end{equation}
To obtain the dark state of the STIRAP process, we solve the time-independent Schr\"odinger equation $H_{0}|\psi \rangle = E |\psi \rangle $ in the R-frame and in the Majorana representation
\begin{equation}
H_{0}^{(R)}(\zeta ) P_{|\psi>^{(R)}}(\zeta ) = E P_{|\psi>^{(R)}}(\zeta ).
\end{equation}
This is a first-order differential equation that can be solved immediately. The solutions are
$P_{|D\rangle^{(R)}} (\zeta ) = \frac{1}{2}e^{i \Theta} \zeta^2 + \frac{1}{2} e^{-i \Theta}$ with 
$E_{D} = 0$ and 
$P_{|n_{\pm}\rangle^{(R)}} (\zeta ) = -\frac{i}{2\sqrt{2}}e^{i\Theta} \zeta^{2} \mp \frac{\sqrt{2}}{2}\zeta + \frac{i}{2\sqrt{2}}e^{-i \Theta}$ with $E_{\pm} = \pm \hbar \Omega /2$. In the case of the dark state, the roots are $\pm i e^{-i \Theta}$, thus corresponding to the points $(\pm \sin \Theta , \pm \cos \Theta , 0)$  in the $xOy$ plane. Transforming to the Dirac representation and back from the R-frame, we recover the usual 
solutions Eq. (\ref{eq:eigenv}) of the eigenvectors-eigenenergies for the on-resonance STIRAP Hamiltonian.

With these preparations, the derivation of the counterdiabatic term is straigthforward. We impose the condition that the time-dependent Schr\"odinger equation $i\hbar (d/dt) |\psi (t)\rangle = [H_{0}(t) + H_{\textrm{cd}}(t)]|\psi (t) \rangle$ admits a solution that follows exactly the dark state $|\psi (t)\rangle=|D(t)\rangle$. In the $R$ frame and in the Majorana representation, this condition becomes
\begin{equation}
i \hbar \frac{d}{dt} P_{|D\rangle^{(R)}}(\zeta ) = H_{\mathrm{cd}}(t)(\zeta ) P_{|D\rangle^{(R)}}(\zeta ),
\end{equation}
which, when using the results above, yields the correct condition $\Omega_{02} = 2 \dot{\Theta }$.

\section{Simulations of adiabatic and superadiabatic dynamics}
\label{sec:simulations}
Here we present the results of simulating the trajectory of the Majorana stars for STIRAP and saSTIRAP under experimentally realistic conditions.

\subsection{Majorana trajectories under H$_0$ and H$_{2\rm{ph}}$}
We simulate the dynamics of a qutrit with transition frequencies $\omega_{01}/2\pi=5.27$ GHz and $\omega_{12}/2\pi=4.82$ GHz. This may be considered as a three-level systems with unequally spaced energy levels with energy level 
spacings $\omega_{01}$ and $\omega_{12}$ and anharmonicity  $\omega_{01}-\omega_{12}$.
This system, initialized in the state $|0\rangle$, is driven resonantly by the STIRAP Hamiltonian with
 $\sigma=35$ ns, $\bar{\Omega}_{01}/2\pi=\bar{\Omega}_{12}/2\pi=45$ MHz, $t_s/\sigma=-1.2$, and it is evolved from 
 $t_i=-182$ ns to $t_f=140$ ns in $1800$ time steps. The driving frequencies of the 
 Gaussian pulses are taken to be resonant to the respective qutrit transition 
 frequencies. We calculate the dynamics of a qutrit and plot the corresponding trajectories on the Majorana 
 sphere as shown in Fig.~\ref{Fig:Maj-STIRAP-H2ph}(a). We see that STIRAP alone is not working perfectly, with cusps appearing along the trajectory, and as a result the final state misses the South Pole. These are due to non-adiabatic transitions, resulting in non-zero populations on state $|1\rangle$ at certain times, as we show in Fig.~\ref{Fig:Maj-STIRAP-H2ph}(b). This is reflected in a non-zero $y$-component of the angular momentum $\langle J_{y}\rangle$ at exactly those times. We have verified numerically for a wide range of STIRAP parameters that non-zero $p_1$ populations of up to 0.05 are obtained when, in the region of maximum transfer, $\dot{\Theta}$ becomes about half the value of $|E_{\pm}|$, i.e. when the standard adiabaticity condition (see Section III B) becomes not so well satisfied.
Instances of occurrence of these non-adiabatic transitions are further characterized using the measure of infidelity, $1-F=|\langle \psi(t)|n+ \rangle|^2 + |\langle \psi(t)|n- \rangle|^2 \approx 2|\langle \psi(t)|n\pm\rangle|^2
$ (the latter approximation is well satisfied numerically), which corresponds to the overlap between the STIRAP driven single-qutrit state and other eigenvectors of $\rm{H}_0$ (Eq.~\ref{eq:eigenv}) as presented in Fig.~\ref{Fig:Maj-STIRAP-H2ph}(b).
Note also that the Majorana representation is very sensitive to all these errors. For example, the final state in Fig.~\ref{Fig:Maj-STIRAP-H2ph}(a) has populations $p_{0}(t_f) = 0.010$, $p_{1}(t_f)= 0.003$, and $p_{2}(t_f) = 0.987$ on the  states $|0\rangle$, $|1\rangle$, and $|2\rangle$ respectively, yet the stars appear clearly distinct and separated from the South Pole. Also the value of the angular momentum which signals the departure from adiabaticity is one order of magitude larger that the population of state $|1\rangle$.
This sensitivity is a general feature of the Majorana representation, which will be visible in all the following figures.

 Further, we examine the counterdiabatic Hamiltonian Eq.~\ref{eq:CD}, which, when acting separately, would drive the system along the dark state. Due to the fact that the direct transition 0 - 2 is forbidden in the transmon, we produce this coupling via a two-photon drive Eq.~\ref{eq:h2ph_lambda}, a process which brings in its own non-idealities.
With the same parameters for the Gaussian pulses, we simulate the two-photon resonant 
 drive $\rm{H}_{2\rm{ph}}$ at different values of the detunings $\Delta$. 
From Eq.~(\ref{eq:h2ph_lambda}) we see that the constant part of the phase, $\phi_{2\rm{ph}}$, shifts the overall plane of the trajectory, while 
time-dependent effects arise from $\Delta t$. This produces a rapidly 
oscillating phase of the drive, yielding  a wiggling trajectory on the Majorana sphere as shown in Fig.~\ref{Fig:Maj-STIRAP-H2ph}(c).
The frequency of this wiggling is $\Delta/2\pi$, which, as expected, is the same as the frequency of the rotating frame.
Thus, for a system with larger (smaller) anharmonicity ($\omega_1-\omega_2$), or for a two-photon resonance pulse which is 
more (less) detuned from the respective $|0\rangle-|1\rangle$ and $|1\rangle-|2\rangle$ transitions, the rate of wiggling is higher (lower),
which further correspond to the rate of wiggling in  $p_1(t)$ as shown in Fig.~\ref{Fig:Maj-STIRAP-H2ph}(d)(i) for detunings $\Delta/2$, $\Delta$ and $2\Delta$.
Interestingly, we observe similar oscillations also in the black curve representing the angular momentum vector.
 Unlike the case of Majorana sphere trajectories (trajectories traversed by the Majorana stars on the sphere), 
 the rate of wiggling in this case relates to the speed of the single-qutrit evolution, 
 which in a given time interval is directly proportional to the pitch of the wiggling black line. Thus, we can observe visually that the 
 evolution is faster in the middle while slower close to the initial and the target states.
Further, the amplitude of the wiggles in the trajectory is more pronounced in the middle of the drive,
when the first excited state $|1\rangle$ gets populated. 
We closely observe these oscillations by plotting the projection of the angular momentum vector in the plane $z=0$, as shown in Fig.~\ref{Fig:Maj-STIRAP-H2ph}(d)(ii-iv), where each point corresponds to the combination $(\langle J_x \rangle, \langle J_y \rangle)$ at a given time, and the consecutive points in time are connected by lines (in yellow). For larger detunings, the rate of wiggling is faster and hence the resulting network is more dense. 
The amplitude of the wiggles is reflected in the area occupied by these projections $(\langle J_x \rangle, \langle J_y \rangle)$, which is largest in 
Fig.~\ref{Fig:Maj-STIRAP-H2ph}(d)(ii), where the detuning of the two-photon drive is relatively small and 
the system possesses lesser anharmonicity -  making it more likely for the first excited state to get populated.
A reverse situation may be seen in Fig.~\ref{Fig:Maj-STIRAP-H2ph}(d)(iv).
Thus, this demonstrates visually the advantage of using a system with large anharmonicity.

%%%%%%%%%% Figure 4
\begin{figure}
 \centering
 \includegraphics[scale=1,keepaspectratio=true]{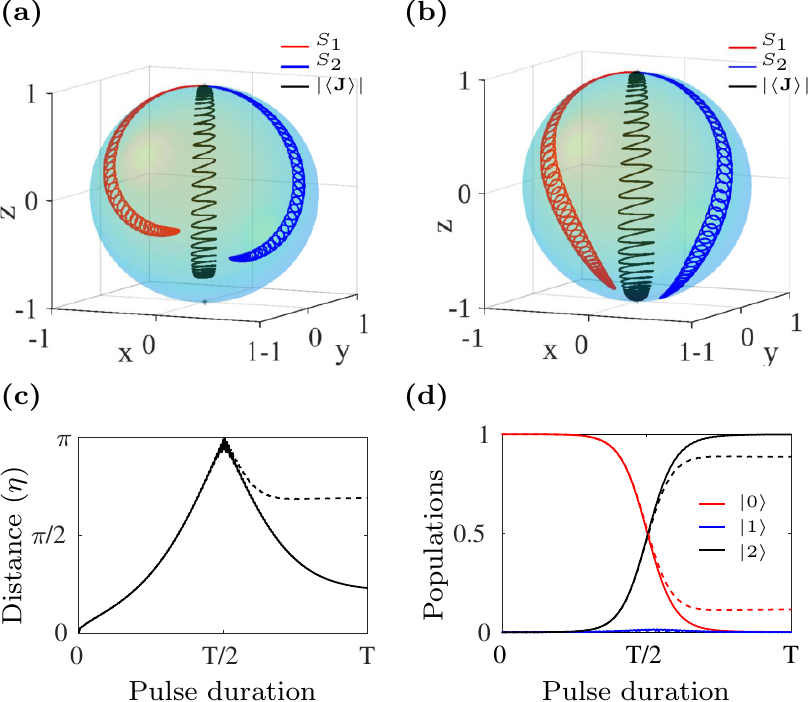}
 % Maj-saSTIRAP.pdf: 0x0 pixel, 300dpi, 0.00x0.00 cm, bb=
 \caption{(a) Majorana trajectories (in red and blue colors) resulting from the two-photon 
 drive with a constant phase $\phi_{\rm 2ph}$ as in Eq.~\ref{eq:h2ph_lambda}. The 
 dynamically phase-corrected Majorana trajectories, with a time-dependent phase $\phi_{\rm 2ph}(t)$, 
 is shown in (b). Plots in (c) and (d) present the corresponding variation of distance between the Majorana 
 stars and the evolution of occupation probabilities with time. The dashed lines are uncorrected values, while the continuous lines include the ac Stark shift correction.
 The parameters are: qutrit transition frequencies, $\omega_{01}/2\pi=5.27$ GHZ, $\omega_{12}/2\pi=4.82$ GHz,
 $\sigma=35$ ns, $\bar{\Omega}_{01}/2\pi=\bar{\Omega}_{12}/2\pi=45$ MHz, $t_s/\sigma=-1.2$, and the state is evolved from 
 $t_i=-182$ ns to $t_f=140$ ns with $T=t_f-t_i$ in $1800$ time steps. Driving frequencies of the 
 Gaussian pulses is taken to be same as that of respective qutrit transition 
 frequencies. \label{Fig:Maj-STIRAP-Stark}}
\end{figure}
The two-photon pulse is also responsible for producing  ac-Stark shifts of the energy levels. These can be compensated 
by using a dynamical phase corrections ~\cite{DiStefano2016,antti-NOTgate-2018,antti-science-2019}.
The correction is applied to the phases of the drives, such that $\phi_{nk}(t)=\phi_{nk}+\int_{\infty}^{t} \epsilon_{nk}(t)dt/\hbar$,
where $n,k=0,1,2$ are the labels of energy levels and
$\epsilon_{nk}(t)$ is the ac Stark shift resulting from the $n-k$ drive at a given time $t$. As shown in Ref.~\cite{antti-NOTgate-2018,antti-science-2019},
respective ac Starks shifts are given by $\epsilon_{01}(t)=\hbar|\Omega_{2\rm{ph}}|^2/\Delta$, 
  $\epsilon_{12}(t)=-5\hbar|\Omega_{2\rm{ph}}|^2/4\Delta$, and $\epsilon_{02}(t)=-\hbar|\Omega_{2\rm{ph}}|^2/4\Delta$. The corresponding dynamic phase corrections are thus obtained as 
$\phi_{01}(t) = \phi_{01} + 2\sqrt{2}\hbar \Theta(t)$, 
$\phi_{12}(t) = \phi_{12} - (5\hbar /\sqrt{2})\Theta(t)$, and $\phi_{02}(t) = \phi_{02} - (\hbar /\sqrt{2})\Theta(t)$.

The trajectories on the Majorana sphere of a qutrit driven by the two-photon resonant pulse  $\rm{H}_{2\rm{ph}}$ and corrected for the 
ac Stark shifts are shown in Fig.~\ref{Fig:Maj-STIRAP-Stark}. 
We simulate the Majorana trajectory with the same parameters of the two-photon resonance as used 
in Fig.~\ref{Fig:Maj-STIRAP-H2ph}(c) and plot the trajectories under H$_{2\rm{ph}}$ with and without the dynamically 
corrected phases compensating the ac Stark shifts in Fig.\ref{Fig:Maj-STIRAP-Stark}(b) and (a) respectively. The relative distance $\eta$ between the Majorana points is shown in Fig.~\ref{Fig:Maj-STIRAP-Stark}(c). As expected, $\eta$
first increases from $0$ to $\pi$ and then decreases from $\pi$ to lower values tending to approach $0$ in the solid black curve with ac Stark shift correction, performing much better than 
that of the dashed curve that represents the distance without the dynamical phase corrections.
The corresponding variation in the populations of the ground state ($p_0$ in red), the first excited state ($p_1$ in blue color) 
and the second excited state ($p_2$ in black) are shown in Fig~\ref{Fig:Maj-STIRAP-Stark}(d), 
where again solid lines represent the respective populations 
with the dynamically modified phases while  dashed lines correspond to the constant phase $\phi_{nk}$.
The advantage of the dynamic phase correction is quite evident from the highly improved population values as 
well as from the Majorana trajectories. However, for simplicity we will not employ it in the following - as we will see, the combined action of $H_{0}$ and $H_{\rm 2ph}$ leads already to high enough transfer fidelities.

%%%%%%%%%%%%%%
\subsection{Majorana trajectory under sa-STIRAP}
%%%%%%%%%% Figure 5
\begin{figure}
 \centering
 \includegraphics[scale=1,keepaspectratio=true]{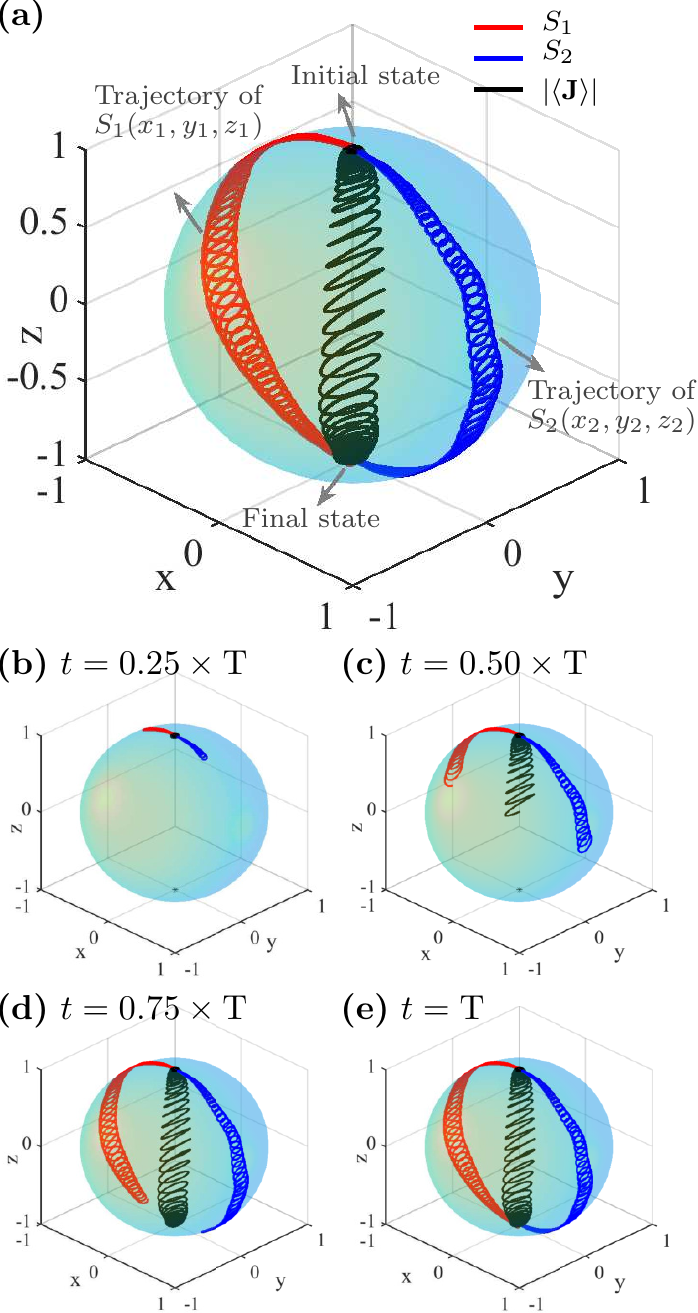}
 % Maj-saSTIRAP.pdf: 0x0 pixel, 300dpi, 0.00x0.00 cm, bb=
 \caption{(a) Majorana trajectories in saSTIRAP, and (b)-(e) their evolution at different moments of time. Parameters: qutrit transition frequencies, $\omega_{01}/2\pi=5.27$ GHz, $\omega_{12}/2\pi=4.82$ GHz,
 $\sigma=35$ ns, $\bar{\Omega}_{01}/2\pi=\bar{\Omega}_{12}/2\pi=45$ MHz, $t_s/\sigma=-1.2$, $\phi_{01}=\phi_{12}=0$, $\phi_{2\rm{ph}}=-\pi/4$, state is evolved from 
 $t_i=-182$ ns to $t_f=140$ ns in $1800$ time steps. The driving frequencies of the 
 Gaussian pulses are the same as that of the  respective qutrit transition 
 frequencies. \label{maj-saSTIRAP}}
\end{figure}

We observe the single-qutrit dynamics under the sa-STIRAP Hamiltonian,
\begin{equation}
 \rm{H}_{\rm saSTIRAP} = \rm{H}_0 + \rm{H}_{2\rm ph},  \label{eq:hsa}
\end{equation}
with the qutrit initialized in the dark state $|D\rangle$ with $\Theta=0$.
$\rm{H}_{\rm saSTIRAP}$ preserves the robustness of the STIRAP Hamiltonian $\rm{H}_0$, 
while concurrently $\rm{H}_{2\rm ph}$ improves the performance, and
precisely returns the expected final state $|D\rangle$ with $\Theta=\pi/2$ in 
an experimentally feasible time. The corresponding trajectory of the qutrit on 
the Majorana sphere is presented in Fig.~\ref{maj-saSTIRAP}.
The simulation is performed with the same set of parameters as before, as specified in the figure caption.
Fig.~\ref{maj-saSTIRAP}(a) shows the full trajectories as the qutrit initialized in $\vert 0\rangle$ (North Pole of the Majorana sphere) evolves under $\rm{H}_{\rm saSTIRAP}$ and tends to approach the second excited 
state $|2\rangle$ (South Pole of the Majorana sphere). Figs.~\ref{maj-saSTIRAP}(b)-(e) present 
the same simulation at four equally spaced values of time. As expected, the dynamics of Majorana stars is slower close to the start and to the end of the process and becomes
faster in the middle.

%%%%%%%%%%
\subsection{STIRAP vs saSTIRAP}
\begin{figure}
 \centering
 \includegraphics[scale=1,keepaspectratio=true]{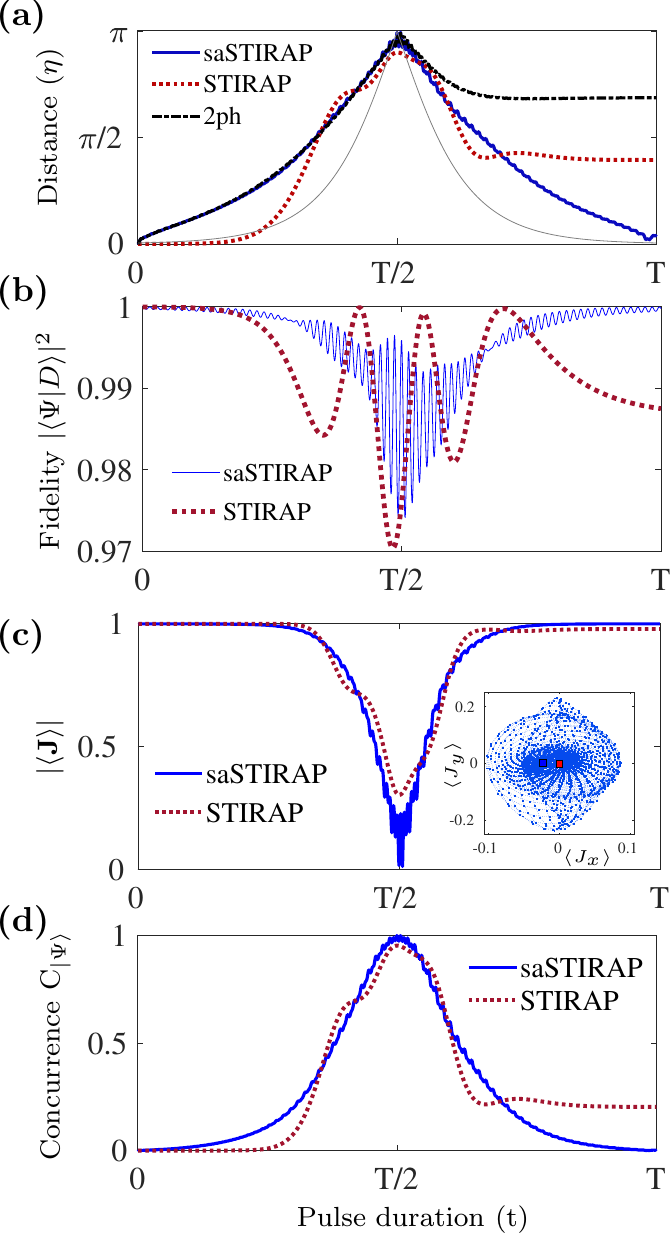}
 % Maj-saSTIRAP.pdf: 0x0 pixel, 300dpi, 0.00x0.00 cm, bb=
 \caption{Process characterization as a function of time. STIRAP is represented with red dotted lines, while saSTIRAP is shown with blue continuous lines. (a) Variation of the angle $\eta$ during STIRAP and saSTIRAP. In addition, the black continuous line represents the two-photon process, while the thin grey line shows a corresponding ideal adiabatic following of the dark state. 
 (b)  Fidelity between the dark state $|D\rangle$ and the time-evolved state of a qutrit $|\Psi(t)\rangle$, under STIRAP and saSTIRAP. (c) Magnitude of the average spin-1 vector for the two processes, showing a dip to nearly zero. The time-evolution of the projections of this vector in the $x0y$-plane is shown in the inset, where the ideally expected state $-|2\rangle$ is represented by a red square and the actual state reached by saSTIRAP is shown by a blue square.
 (d) Concurrence in the symmetrized-state representation
showing a peak to nearly 1 in the middle.
 \label{Fig:theta_all}} \label{pop-saSTIRAP}
\end{figure}

A thorough comparison between the STIRAP and saSTIRAP processes (with same parameters as in Fig.~\ref{maj-saSTIRAP}) is carried out in Fig.~\ref{Fig:theta_all}. We start with the time-dependence of the distance $\eta$
between the Majorana stars of a qutrit starting from $|0\rangle$ as it evolves under H$_0$, H$_{2\rm{ph}}$ and H$_{\rm saSTIRAP}$. The 
plots are shown in Fig.~\ref{Fig:theta_all}(a), where the dashed red curve presents the variation of $\eta$ under H$_0$, the black dot-dashed curve corresponds to $H_{2\rm{ph}}$, and the blue continuous curve 
corresponds to the evolution under the H$_{\rm saSTIRAP}$ Hamiltonian. The continuous blue curve of saSTIRAP clearly gives the best result. For comparison, we 
present with a grey continuous line an ideal saSTIRAP process, where the dark state is followed perfectly.

Further, we compare systematically the imperfections resulting from practically feasible STIRAP and saSTIRAP drives
with respect to an ideal adiabatic dynamics, which strictly follows the dark state $|D\rangle$. At various values of the time $t$ we calculate the fidelity $F(t)=|\langle \Psi(t) | D \rangle|^2$, where $| \Psi(t)\rangle$ represents either the STIRAP or saSTIRAP -evolved state at an arbitrary time $t$, see also Eq. (\ref{eq:fidelity}). The plots  presenting the variation of fidelity $F(t)$ 
are shown in Fig.~\ref{Fig:theta_all}(b). A closer look at the fidelity curves reveals that, despite the remaining small oscillations, the state under saSTIRAP is closer to the dark state, while under STIRAP it is relatively far from the desired behaviour. The oscillations of fidelity seen in STIRAP coincide with the appearance of cusp-like structures in the Majorana trajectories shown in Fig.~\ref{Fig:Maj-STIRAP-H2ph}(a). The magnitude of the spin vector $|\langle\mathbf{J}\rangle|$ is another way  to characterize the two processes, as shown in Fig.~\ref{Fig:theta_all}(c). Finally, the concurrence in the symmetrized state picture is obtained in \ref{Fig:theta_all}(d). Corresponding to the dip in $|\langle\mathbf{J}\rangle|$ in the middle of the evolution, the concurrence develops a peak. The two spin-1/2 particles become maximally entangled, and as a result the partial density matrix is almost maximally mixed $\mathbb{I}_{2}/2$ with the $r$-parameter close to zero. 
%Overall, we find that the Majorana representation is very sensitive towards the imperfections of the processes, and as such can be used as a tool for evaluating the errors in gates  based on shortcuts to adiabaticity. 

%%%%%%%%%%% Figure 6

\subsection{Majorana representation of mixed states}

In real experiments, due to decoherence, the states are mixed. However, the problem of extending the Majorana picture to mixed states does not have a straightforward solution. Several proposals have been put forward, for example using multiaxial representations~\cite{sirsi-pra-2017}, probability simplexes~\cite{lopez-entropy-2018}, elimination methods~\cite{karol-book-2013}, and projective spaces of polynomials~\cite{Serrano-pra-2020}. Here we address this problem with the goal of finding a solution that is of closest relevance to experimental practice. In experiments, the usual situation is that decoherence exists but measures are taken to reduce it as much as possible. The manipulation of qubits and qutrits by applying quantum gates is done on shorter times than the decoherence times. Thus, the system approximately follows the intended pure state, and the mixedness can be regarded as a perturbative effect. To account for this situation, we consider a mixed qutrit state $\rho$ and perform a spectral decomposition
\begin{equation}
\rho = \sum_{i=d,e,f}\lambda_{i}|\chi_{i}\rangle\langle \chi_{i}| \label{Eq:rho_mixed}
\end{equation}
where $\lambda_{i} \geq 0$. This means that we regard our mixed state as a statistical ensemble of states $|\chi \rangle$ occuring with probabilities $\lambda_{i}$. For low levels of decoherence, one of  these probabilities (say $\lambda_{d}$) will be the largest, and it will correspond to the state along which we intend to drive the system  (the dark state in our case). Then, we represent this state, together with the remaining two states, as stars on three Majorana spheres, with radiuses $\lambda_{d}, \lambda_{e}$, and $\lambda_{f}$ respectively. For example, in the case of a spin-1/2 (qubit) mixed state, with standard representation $\mathbb{I}/2 + \mathbf{r}\boldsymbol{\sigma}/2$ where $\mathbf{r} = (r \sin\theta \cos \varphi , r \sin \theta \sin \varphi , r \cos \theta )$ we obtain $\lambda_{e} = (1+ r)/2$, $\lambda_{f} = (1-r)/2$, and $|\chi_{e}\rangle = \cos (\theta /2 ) |0\rangle + \sin (\theta /2) \exp(i \varphi )|1\rangle$, $|\chi_{f}\rangle = \sin (\theta /2 ) |0\rangle - \cos (\theta /2) \exp(i \varphi )|1\rangle$. The representation produces two points in opposite directions on two spheres with subunit radiuses $\lambda_{e}$ and $\lambda_{f}$. Eigenvalue degeneracy (maximally mixed state) occur when the two spheres overlap, in which case we do not know which point to associate to which sphere and any two orthogonal vectors could be used to represent the state; this corresponds, in the standard representation, to the Bloch sphere shrinking to a point.

%%%%%%%%%% Figure 7
\begin{figure}
 \centering
 \includegraphics[scale=1,keepaspectratio=true]{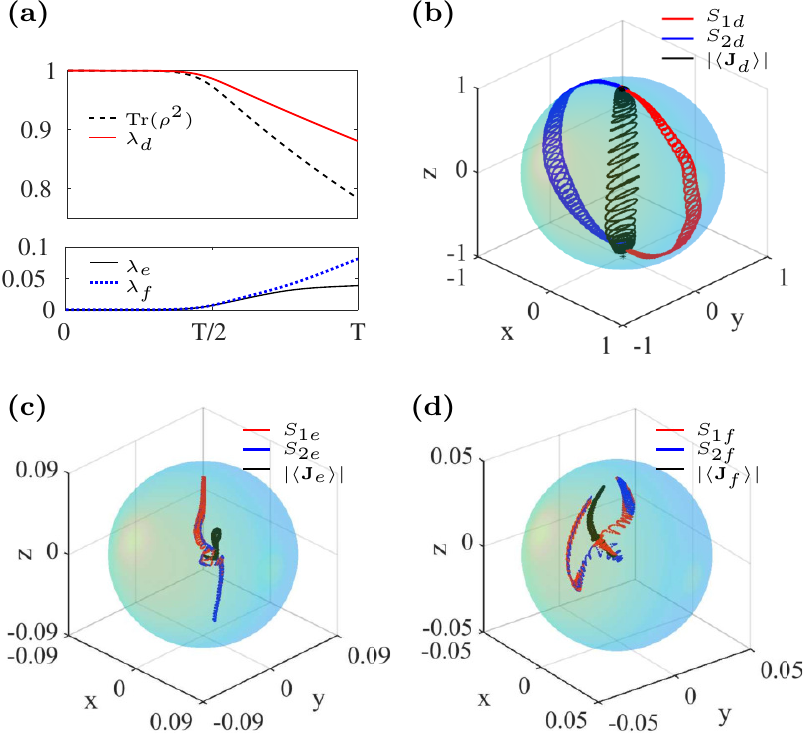}
 % Maj-STIRAP-mixed1.pdf: 0x0 pixel, 300dpi, 0.00x0.00 cm, bb=
 \caption{Majorana representation of the eigenvectors of the single-qutrit state $\rho(t)$ driven by the saSTIRAP
 Hamiltonian in a decohering environment. The parameters of the drive are the same as in Fig. 6.
 (a) Time-variation of the purity $\rm{Tr}[\rho(t)^2]$ and the eigenvalues $\lambda_i(t)$ with  $i\in\{d,e,f\}$ of 
 $\rho(t)$. (b) saSTIRAP driven trajectories of 
 the Majorana stars of $|\chi_d(t)\rangle$ shown on a sphere of radius $\lambda_d(t)$. This radius is unity in the beginning of the process, and gradually decreases with time. The black asterix at the bottom of the sphere marks the South Pole.  The dynamics of the Majorana stars of $|\chi_d(t)\rangle$ clearly are very close 
 to that of the dark state.
 (c),(d) present the corresponding trajectories of states $|\chi_e(t)\rangle$ and $|\chi_f(t)\rangle$
 respectively. The spherical surfaces shown correspond to $\lambda_{e,f}(t)$ at the end of the process.}
 \label{fig:mixed1}
\end{figure}

To illustrate this concept, we consider a qutrit initialized in the ground state ($|0\rangle$),
and driven by the saSTIRAP Hamiltonian in the presence of decoherence. The evolution
of the system is governed by the Lindblad master equation, $\dot{\rho}=-i/\hbar [\rm{H},\rho]+\mathcal{L}[\rho]$, where 
$\mathcal{L}[\rho]= \Gamma_{21}\rho_{22}(|1\rangle\langle 1| - |2\rangle\langle 2|) + \Gamma_{10}\rho_{11}(|0\rangle\langle 0| - 
 |1\rangle\langle 1|) - \sum_{j,k\in{0,1,2}, j\neq k}\gamma_{jk}\rho_{jk}|j\rangle\langle k|$ 
with $\gamma_{10}= \Gamma_{10}/2 + \Gamma_{10}^{\phi}$, $\gamma_{20} = \Gamma_{21}/2 + \Gamma_{20}^{\phi}$ and $\gamma_{21} = (\Gamma_{10}
+ \Gamma_{21})/2 + \Gamma_{21}^{\phi}$
 is the Lindblad qutrit superoperator \cite{kumar-nature-2016}.
We simulate the evolution of this equation with relaxation rates $\Gamma_{10}=0.5$ MHz and $\Gamma_{21}=0.71$ MHz and with pure dephasing rates 
$\Gamma_{10}^{\phi}=0.4$ MHz, $\Gamma_{21}^{\phi}=0.56$ MHz, and $\Gamma_{20}^{\phi}=0.96$ MHz, which correspond to realistic experimental values for transmons.
Due to the interaction between the system and the environment, our single-qutrit state becomes mixed. Applying the spectral decomposition Eq.~\ref{Eq:rho_mixed}, we obtain 
$\rho(t)$ at any time $t$ as a linear combination of mutually orthogonal states $|\chi_{i}(t)\rangle \langle \chi_{i}(t)|$. The time-variation of the 
weights $\lambda_i$s of the various component vectors $|\chi_{i}\rangle$s are shown in Fig.~\ref{fig:mixed1}(a), 
together with the measure of purity $\rm{Tr}[\rho(t)^2]=\lambda_d(t)^2+\lambda_e(t)^2+\lambda_f(t)^2$. The fidelity with which the final state matches the ideally expected 
saSTIRAP dynamics is obtained using $\mathcal{F}(t)=|\langle D(t) |\rho(t)|D(t)\rangle|$, and is found out to be $0.88$.
Further, we obtain the Majorana representation of the eigenvectors ($|\chi_d (t)\rangle$, $|\chi_e (t)\rangle$, $|\chi_f (t)\rangle$). Each of these eigenvectors has two Majorana roots, $\mathbf{S}_{1i}$ and $\mathbf{S}_{2i}$, which correspond to pairs
of Majorana stars on spheres with radiuses  
$\lambda_d(t)$, $\lambda_e(t)$, $\lambda_f(t)$. For clarity, we plot these starts as separate pictures, see Fig.~\ref{fig:mixed1}(b,c,d). 
From  Fig.~\ref{fig:mixed1}(b) one can see that the state $|\chi_d\rangle$, which has the largest statistical weight in $\rho$, follows closely the dynamics of the dark state $|D\rangle$. Under the saSTIRAP drive, this pair
of Majorana stars starts from the North Pole at $t=0$ ($\lambda_d(0)=1$) and would ideally move symmetrically towards the South Pole, along the plane $y=0$. However, as the decoherence sets in,
$\lambda_d<1$, and trajectories of the Majorana stars go inside the radius-1 sphere, see Fig.~\ref{fig:mixed1}(b).
 As $\lambda_d$ decreases, $\lambda_{e(f)}$ become non-zero, and correspondingly, trajectories of the Majorana stars representing the 
dynamics of states $|\chi_{e}\rangle$ and $|\chi_{f}\rangle$
appear to emerge from the center of the Majorana mixed sphere with radiuses $\lambda_{e}(t)$ and  $\lambda_{f}(t)$
as is clearly evident from Fig.~\ref{fig:mixed1} c),d).
We have also performed simulations for only the STIRAP drive in the same decohering environment, and we have obtained similar results. Also in this case the statistical mixture of $\rho(t)$ consist predominantly of the dark state $|D(t)\rangle$.  A final state fidelity (at $t=T$) of $0.868$ is reached in this case. 

Furthermore, our geometric representation for the angular momentum can be extended to mixed states in a straightforward way. The Majorana stars $\mathbf{S}_{1i}$, $\mathbf{S}_{2i}$ are now defined on spheres with radiuses $\lambda_{i}$. Correspondingly, the bisector of the angle $\eta_i=\cos^{-1}(\mathbf{S}_{1i}\cdot\mathbf{S}_{2i})$ is $\mathbf{OO'}_{i}=(\mathbf{\rm{\mathbf{S}}}_{1i}+\mathbf{\rm{\mathbf{S}}}_{2i})/2$ and the total averaged angular momentum $\langle \rm{\mathbf{J}} \rangle = \sum_{i=d,e,f}\langle \rm{\mathbf{J}_{i}} \rangle$
can be obtained geometrically via
\begin{equation}
\langle \rm{\mathbf{J}_{i}} \rangle=\frac{2\lambda_i\mathbf{OO'}_{i}}{\lambda_i^2+|\mathbf{OO'}_{i}|^2},
\end{equation}
which generalizes Eq. (\ref{eq:J1}) to mixed states. The $\langle \rm{\mathbf{J}_{i}}\rangle$ vectors are shown separately in Fig. \ref{fig:mixed1} b)c)d) with black lines.

To summarize, we introduced a geometrical representation that require the specification of three pairs of Majorana stars and three radiuses that add up to unity 
in order to completely and uniquely determine the single-qutrit state $\rho$ in the 
three-dimensional Hilbert space. This procedure can be generalized to larger dimensions in a straightforward way.

We have also performed quantum process tomography~\cite{chuang-jmod-1997} to characterize the STIRAP and saSTIRAP protocols for the case of open-system qutrit. We prepare our qutrit in nine initial states ($\rho_i$, $i \in \{1,9\}\} $), that form the complete basis for single-qutrit density matrices. Each of these initial states are allowed to evolve under the same process, which is to be characterised. We simulate the evolution of the system under the STIRAP and saSTIRAP dynamics, both in the presence and absence of the system-environment interaction, governed by the Lindblad master equation and obtain the corresponding process matrices. For both STIRAP and saSTIRAP, we find that the process matrices for the case of decoherence with the parameters above match with the process matrices for the no-decoherence case with a fidelity of $\approx 0.77$ (corresponding to a trace distance of $\approx 0.25$).

To conclude this subsection, we see that even in the presence of decoherence the Majorana representation can be extended to describe the STIRAP and saSTIRAP dynamics.

\subsection{Discussion}

Although the simulations presented in this section were done with parameters specific to one experimental setup - a three-level superconducting circuit driven by three microwaves - we emphasize that the results are more general. Indeed, processes such as STIRAP and most recently also saSTIRAP have  been  implemented on  a remarkable large array of experimental platforms \cite{Bergmann_2019, Torrontegui13}. For example, while in our analysis the introduction of the symmetrized-state representation appears as an abstract mathematical construct, leading to the concurrence result plotted in Fig. 6 c), there are well-established experimental cases when STIRAP does involve two physically distinct objects. One example is the qubit-cavity system which appears in both atomic physics \cite{PhysRevLett.89.067901} and circuit QED \cite{Palmer2017}. 
Another experimental realization  is the transfer of population between two spatially distant qubits via a lossy transmission line \cite{Cleland2020}. Stopping these processes mid-way 
\cite{antti-NOTgate-2018} results in a maximally entangled state between the cavity and the atom (qubit), as predicted by our approach. Another experimental platform is ultracold atomic gases. In the case of bosonic atoms, the spin-coherent state introduced in Section II is a Bose-Einstein condensate of $N=2j$ particles, with macroscopic wavefunction
\begin{equation}
|j,\hat{\mathbf{n}}\rangle = \frac{1}{\sqrt{N!}}\left[ \cos \frac{\theta}{2}a^{\dag}  + \sin \frac{\theta}{2}e^{i \varphi}b^{\dag} \right]^N|\emptyset\rangle ,
\end{equation}
where $|j,m\rangle = (1/\sqrt{(j+m)!(j-m)!})(a^{\dag})^{j+m}(b^{\dag})^{j-m}|\emptyset \rangle$, $|\emptyset \rangle$ is the vacuum state,
and the angular momentum can be written in the Schwinger representation as $J_{x} = (a^{\dag}b + b^{\dag}a)/2$, $J_{y} = (a^{\dag}b - b^{\dag} a)/2$, and $J_{z} = (a^{\dag}a - b^{\dag}b)/2$. The operators $a$ and $b$ typically correspond either to internal atomic states or to the localized states of a two-well potential \cite{RevModPhys.73.307,Paraoanu_2001, PhysRevA.66.013609}. The case $N=2$ is relevant for the formation of ultracold molecules by STIRAP \cite{PhysRevLett.98.043201, Danzl1062,Ni231} and has further attracted a lot of interest due to the experimental possibilities of realizing, controling, and observing only two atoms in optical lattice \cite{Twoatoms,PhysRevA.79.052118, PINTO2009581,PhysRevA.81.043609, PhysRevLett.109.116405}. We note that, for this system, the signature of Bose-Einstein condensation is zero concurrence, while finite concurrence signals the appearance of a fragmented state \cite{PhysRevA.59.3868,
PhysRevA.74.033612, PhysRevA.77.041605}.

%%%%%%%%%%%%%%
\section{ Conclusions \label{sec-conclusions}}
We have investigated the Majorana representation of the evolution under STIRAP and superadiabatic STIRAP. We have shown that the dark states are represented by two stars on the circle defined by the intersection of the $xOz$ plane with the Majorana sphere. We have also introduced a representation of the spin-1 average vector, which evolves along the $Oz$ axis, and we have shown 
how its rate of change in the three-dimensional space is a measure of a state change of the qutrit in the Hilbert space. This vector can be used to characterize the degree of non-adiabaticity  of the processes. The representation puts clearly in evidence the role of the counterdiabatic drive in saSTIRAP that corrects the deviations of the trajectory from the adiabatic path, and as such it offers a sensitive visual diagnosis tool for errors caused by non-adiabaticity and ac Stark shifts. Interestingly, this drive can be derived by using the Majorana polynomial.
We have done an in-depth analysis of the Majorana trajectories resulting from STIRAP and saSTIRAP with and without the dynamically corrected phases used to compensate for the ac Stark shifts, and also for evolutions that include decoherence. We have also analyzed the effectiveness of the STIRAP and saSTIRAP processes via the distance between the Majorana stars.
%%%%%%%%%%%%%%
%
%
%
\section{ Acknowledgements \label{sec-ack}}
We acknowledge support from the Foundational Questions Institute Fund (FQXi) via the grant no. FQXi-IAF19-06, 
%from the FQXi project ``Exploring the fundamental limits set by thermodynamics in the quantum regime'', 
from the European Union Horizon 2020 research and innovation  program  (grant  agreement  no. 862644, FET  Open  QUARTET), and from the Academy of Finland through the RADDESS programme (project no. 328193) and the “Finnish
Center of Excellence in Quantum Technology QTF” (project 312296).
%%%%%%%%%%%%%%
 %\bibliography{majorana-stirap}
%merlin.mbs apsrev4-1.bst 2010-07-25 4.21a (PWD, AO, DPC) hacked
%Control: key (0)
%Control: author (8) initials jnrlst
%Control: editor formatted (1) identically to author
%Control: production of article title (-1) disabled
%Control: page (0) single
%Control: year (1) truncated
%Control: production of eprint (0) enabled
%

%%%%%%%%%%%%%%%%%%%%%%%%
\end{document}